\documentclass[aps,prl,reprint,twocolumn,groupedaddress]{revtex4}
\usepackage{graphicx,color}% Include figure files
\usepackage{xtab,afterpage}
\usepackage{lipsum}
\usepackage{bm}% bold math
\usepackage{amssymb,amsmath,amsfonts, mathrsfs}
\newcommand{\be}{\begin{equation}}
\newcommand{\ee}{\end{equation}}
\newcommand{\ba}{\begin{array}}
\newcommand{\ea}{\end{array}}
\newcommand{\bqa}{\begin{eqnarray}}
\newcommand{\eqa}{\end{eqnarray}}

\begin{document}

%\preprint{}

%\title{Analysis of the pole position of $\psi(2^3D_1)$ by a coupled-channel scheme}
\title{A new look at $\psi(4160)$ and $\psi(4230)$}


%% \affiliation can be followed by \email, \homepage, \thanks as well.
\author{Zhi-Yong Zhou}
\email[]{zhouzhy@seu.edu.cn}
%\thanks{}
\affiliation{School of Physics, Southeast University, Nanjing 211189,
P.~R.~China}
\author{Chun-Yong Li}
%\thanks{}
\affiliation{School of Physics, Southeast University, Nanjing 211189,
P.~R.~China}
%\affiliation{School of Physics, Southeast University, Nanjing 211189,
%P.~R.~China}
\author{Zhiguang Xiao}
\email[]{xiaozg@scu.edu.cn}
%\thanks{}
\affiliation{Physics school, SiChuan University, Chendu, SiChuan, China}

\date{\today}

\begin{abstract}
By simultaneously analyzing the cross section data of $e^+e^-\rightarrow D\bar D, D\bar D^*, D^*\bar D^*, D\bar D\pi$  in a coupled-channel scheme with unitarity, we found that, in contrast to the conventional wisdom, the pole of $\psi(2^3D_1)$ might be located at  about $\sqrt{s}=4222-32i\mathrm{MeV}$.   This observation implies a possibility that the two resonances, dubbed the $\psi(4160)$ and  $\psi(4230)$ in the PDG table now, might be the same $\psi(2^3D_1)$ state. Such a suggestion could provide more insight to our understanding the enigmatic decay properties of  $\psi(4160)$ and $\psi(4230)$. Furthermore, this coupled-channel scheme could be applied to study other phenomena with several interfering resonances.
\end{abstract}

%\pacs{12.39.Jh, 13.25.Gv, 13.75.Lb, 11.55.Fv}
%\keywords{}

\maketitle
\emph{Introduction}: In recent years, the Particle Data Group (PDG)
has updated the parameters of vector charmonium-like states,
$\psi(4160)$ and $\psi(4260)$,  thanks to the improved data analysis techniques and
 the  higher statistics of the data~\cite{ParticleDataGroup:2022pth}. Although the
updated mass and width parameters of these two states are closer to
each other, they are commonly regarded as different states with the
same quantum numbers whose underlying nature remains elusive. The
$\psi(4160)$ was first observed by DASP several decades ago as a
resonance structure around 4159 MeV in the experimentally measured $R$
values~\cite{DASP:1978dns,Seth:2004py}, and it is regarded as a
$2^3D_1$ $c\bar{c}$ state due to its consistency with the predictions
of the quark potential model~\cite{Godfrey:1985xj}. However, the analysis
done by BES reported higher Breit-Wigner~(BW) mass and width
parameters for $\psi(4160)$ as  $M=4191.7\pm 6.5$ MeV and
$\Gamma=71.8\pm 12.3$MeV~\cite{BES:2007zwq}. LHCb also obtained a
similar result by  analyzing the data of  $B^+\rightarrow
K^+\mu^+\mu^-$~\cite{LHCb:2013ywr}. On the other hand, the
$\psi(4260)$ was originally observed as a broad structure by BaBar and
Belle in the $e^+e^-\rightarrow \pi^+\pi^-J/\psi$
process~\cite{BaBar:2005hhc,Belle:2007dxy}, and its possible exotic nature as a
molecular
state~\cite{Ding:2008gr,Li:2013bca,Dai:2012pb,Cleven:2013mka}, a hybrid
state~\cite{Zhu:2005hp,Kou:2005gt,Close:2005iz}, or a $c\bar{c}(4S)$
charmonium state~\cite{Llanes-Estrada:2005qvr} has been suggested. BES
has revealed the non-trivial asymmetry of the broad structure by
 high-statistics data and identified the lower peak of the broad
structure in the $\pi^+\pi^-J/\psi$ mass spectrum as $\psi(4230)$,
whose measured BW mass and width parameters are $M=4222.0\pm 3.1\pm
1.4$ MeV and $\Gamma=44.1\pm 4.3\pm 2.0$ MeV respectively~\cite{BESIII:2016bnd}. Analyses
based on the data in other hidden-charm channels
 exhibit some
variations~\cite{BESIII:2019qvy,BESIII:2019gjc,BESIII:2021njb}. The
$\psi(4160)$ and $\psi(4230)$ have the same quantum numbers with  mass difference of about 30 MeV  but can
hardly be accommodated in the quark model at the same time~\cite{Godfrey:1985xj,Eichten:1979ms,zhao2023mass}.
Furthermore, while the $\psi(4160)$ appears in the open-charm channels, it
is absent in the hidden-charm channels, and the decay channels of
$\psi(4230)$ listed in the PDG table are mostly hidden-charm channels
except in $\pi D^{(*)}\bar D^*$ channels
~\cite{BESIII:2018iea,BESIII:2023cmv}. These observations contribute
to the enigmatic nature of these two states.

Although multiple experimental analyses have suggested the existence of a resonance, $\psi(4160)$, below 4200 MeV, contributing to the shoulder of the cross section structure, it is crucial to be cautious while applying the
Breit-Wigner (BW) parameterization in analyzing the
data with interfering resonances. This is because the use of the BW parameterization introduces certain well-known uncertainties, which must be carefully examined. For instance, when multiple resonances with the same quantum numbers are parameterized in the same energy region, there can be multiple solutions with similar fitting quality, known as the multi-solution ambiguity \cite{Zhu:2011ha, Bukin:2007kx, Han:2018wbo, Bai:2019jrb}. Additionally, when the scattering amplitude is represented by several interfering BW resonances, careful treatment of the interference effect can improve the fit results, as demonstrated in Ref. \cite{BES:2007zwq}. The existence of additional uncertainties cannot be definitively dismissed without considering the constraints of unitarity. As a result, a coupled-channel framework that upholds the principles of coupled-channel unitarity is vitally necessary for the accurate determination of resonance parameters, specifically the pole positions through a proper analytical continuation. Actually, the inclusion of coupled channels has proved to be crucial to the understanding of the properties of many exotic states~\cite{Eichten:1979ms, Tornqvist:1995kr, Li:2009pw, Cao:2014vca,
Limphirat:2013jga, Achasov:2012ss, Zhang:2009gy, Segovia:2011zza,
Uglov:2016orr}.

In this paper, we present a coupled-channel scheme that extends the Friedrichs-Lee model to simultaneously analyze cross-section data from $e^+e^-\rightarrow D\bar D, D\bar D^*, D^*\bar D^*$, and $D\bar D\pi$ processes, as measured by Belle~\cite{Belle:2007qxm, Belle:2006hvs, Belle:2007xvy, Belle:2009dus, Belle:2017grj}. By satisfying the coupled-channel unitarity, our scheme extracts the poles of $\psi(2^3D_1)$ on the nearest unphysical Riemann sheet (RS) through analytical continuation of the amplitude to the complex energy plane. Our results suggest an alternative solution to resolving the enigma surrounding the $\psi(4160)$ and $\psi(4230)$ states that does not rely on an exotic origin for these states. We anticipate that this analysis will provide further insights into the nature of this states.

\emph{The Scheme}: The Friedrichs-Lee
model~\cite{Friedrichs:1948,Lee:1954iq} could be generalized to the
form with multiple discrete states and multiple continuous states
after partial wave decomposition,
dubbed the GFL scheme here~\cite{Xiao:2023,Xiao:2016mon}. The free
Hamiltonian $H_0$ with multiple discrete eigenstates and multiple
continuum states can be represented as
\begin{align}
H_0=&\sum_{\rho=1}^D M_\rho|\rho\rangle\langle
\rho|+\sum_{i=1}^C \int_{a_i}^\infty \mathrm d \omega
\,E|E;i\rangle\langle E;i|,
\end{align}
where $a_i$ is the $i$-th threshold energy. In the followng, Greek
letter indices such as $\rho$ denotes  a discrete state and the Roman letter denotes a continuum state.
The interaction between the $\rho$-th discrete state and the $i$-th continuum state is supposed to be described by a general coupling function $f_{\rho i}(E)$, thus the interaction Hamiltonian $V$ reads
\bqa
V=\sum_{\rho=1}^D\sum_{i=1}^C  \int_{a_i}^\infty\mathrm d E
[f^*_{\rho i}(E)|\rho\rangle\langle E;i|+f_{\rho i}(E)|E;i\rangle
\langle \rho|] .
\eqa
The eigenvalue problem for this Hamiltonian could be rigorously solved, and the partial-wave scattering amplitude of the $i$-th continuum to the $j$-th continuum could be written down as
\begin{align}\label{PWSmatrix}
&S_{i,j}=\delta_{ij}-2\pi i \sum_{\rho,\lambda} f^*_{\rho j}(E) [\eta^{-1}]_{\rho\lambda}    f_{\lambda i}(E),
\end{align}
where $[\eta^{-1}]$ represent the resolvent matrix and the element of its inverse matrix reads
\bqa\label{etafunction}
[\eta]_{\rho\lambda}\equiv(E-M_\rho)\delta_{\rho\lambda}-\sum_{i=1}^C \int_{a_i}^\infty
\mathrm d E' \frac{f_{\rho i}^*(E')f_{\lambda i}(E')}{(E-E')}.
\eqa
It is straightforward to confirm that the scattering matrix
satisfies the partial-wave unitarity, i.e., $SS^\dag=1$. In this
study, we employ the Quark Pair Creation (QPC)
model~\cite{Micu:1968mk,Blundell:1995ev} to describe the coupling
vertex function $f_{\rho}(E)$ between the charmonium bare state and
the open-charm continuum state. In this model, the strong coupling
occurs through the creation of a quark-antiquark pair with vacuum
quantum number $J^{PC}=0^{++}$ and the interaction can be derived from
the quantum field theory Hamiltonian
\bqa
H_I=g\int d^3 x\bar{\psi}(x)\psi(x), \ \ \ t=0,
\eqa
where $\psi(x)$ is a Dirac field operator~\cite{Ackleh:1996yt}. Following the standard procedures, one can obtain the coupling vertex function as explicitly derived in refs.~\cite{Zhou:2017txt,Blundell:1995ev}.

This representation of the $S$ matrix satisfies the required
analyticity by the coupled-channel unitarity as it is  continued to
the complex energy plane. With each unitarity cut doubling the number
of RS, the amplitude with $n$ thresholds has $2^n$ RSs. However, the
most important RSs are those directly connected to the physical region
since the poles on those sheets  near the physical region will
significantly influence the experiment observables. In this study, we
refer to the RS connected to the physical region between the $i$-th
and higher thresholds as the $(i+1)$-th sheet. The last sheet of
interest is the one linked to the physical region above the highest
threshold. Other unphysical sheets are distant from the physical
region, and the poles on these sheets have a negligible impact on the
analysis, so we will not discuss them.

Consequently, since the scattering amplitude in eq.(\ref{PWSmatrix}) is in a matrix form, the poles of the scattering amplitude on the $N$-th RS will correspond to the zeros of the determinant of the $\eta$ matrix on the same sheet. That means, the poles' position could be extracted by solving
\bqa\label{deteta}
\mathrm{Det}[\eta^{(N)}(z_0)]=0
\eqa
on the complex energy plane, in which $z_0=M-\frac{\Gamma}{2}i$ where $M$ and $\Gamma$ represent the pole mass and width respectively.

%Consequently, the $\eta$ function matrix on the $N$-th sheet~($N=II, III, \cdots, n+1$) is defined as
%\bqa\label{etaN}
%[\eta^{(N)}]_{\rho\lambda}(z)\equiv[\eta^{(I)}]_{\rho\lambda}(z)-2\pi
%i \sum_{j=1}^{N-1} {\red f_{\rho j}^*(z)f_{\lambda j}(z)}.
%\eqa
%Since the scattering amplitude in eq.(\ref{PWSmatrix}) is in a matrix form, the poles of the scattering amplitude on the $N$-th RS will correspond to the zeros of the determinant of the $\eta$ matrix on the same sheet. That means, the poles' position could be extracted by solving
%\bqa
%\mathrm{Det}[\eta^{(N)}(z_0)]=0
%\eqa
%on the complex energy plane, in which $z_0=M-\frac{\Gamma}{2}i$ where $M$ and $\Gamma$ represent the pole mass and width respectively.

Apart from the strong interactions between the hadronic continuum
states and the resonance, the electroweak interactions of $\psi$ to the $e^+e^-$ continuum $\psi_\rho\rightarrow e^+e^-$
could be studied generally as
\bqa
\mathcal{M}(\psi_\rho\rightarrow e^+e^-)=\frac{eg_{e\rho}}{M_\rho^2}(\bar u\gamma_\mu v)\epsilon^\mu,
\eqa
where $u$ and $v$ are the wave function of the electron and positron, and $\epsilon^\mu$ is the polarization vector of the $\psi_\rho$ meson. $M_\rho$ is its bare mass and $g_{e\rho}$  is the  coupling constant.  The electroweak interaction strength is much weaker than the strong interaction, so it could be studied in a perturbative manner, and the cross section of the annihilation process $e^+e^-\rightarrow (continuum)_i$ is written down as
\bqa\label{crosssection}
\sigma_i(E)=\frac{32\pi^5\alpha}{E^5}|\sum_{\rho\lambda}g_{e\rho}[\eta^{-1}]_{\rho\lambda}f_{\lambda i}|^2,
\eqa
where $\alpha=\frac{e^2}{4\pi}=\frac{1}{137}$ and $e$ is the electron charge. The integral term associated with the $e^+e^-$ channel in the inverse resolvent matrix $[\eta]$ is negligible and does not need to be considered.

\emph{Numerical analysis}: We first give an account of  the continuum states,
the discrete states and their coupling functions in this analysis.
With this coupled-channel scheme, we simultaneously fit the updated
high-statistics  cross section data of the $e^+e^-\rightarrow D\bar
D$~\cite{Belle:2007qxm}, $D\bar D^{*}$~\cite{Belle:2017grj},
$D^{*}\bar D^{*}$~\cite{Belle:2017grj}, $D\bar
D\pi$~\cite{Belle:2007xvy} processes measured by Belle in 2007 or
2017.  The previously published Belle data~\cite{Belle:2006hvs} and data obtained from other experiments such as BABAR~\cite{BaBar:2006qlj, BaBar:2009elc}, CLEO~\cite{CLEO:2008ojp}, and BES~\cite{BESIII:2021yvc} are consistent with the Belle data used in our analysis. However, due to the limited range of data or insufficient statistics, these data sets were not included in our fit. Apart from the
electroweak $e^+e^-$ channel, the strong interaction typically
involves four types of hadronic channels: $D\bar D$, $D\bar D^*$,
$D^*\bar D^*$, and $D\bar D\pi$. To simplify the numerical analysis of
the $D^{(*)}\bar D^{(*)}$ channels, we assume that the isospin symmetry is
precisely conserved, and the thresholds for the charged and neutral
modes are degenerate. The average values of cross section data and
their uncertainties of $e^+e^-\rightarrow D^+ D^-$ and $D_0\bar D_0$
are used in the fit~\cite{Belle:2007qxm}. For the $D^*\bar D^*$
channels, there could be three different combinations of  the total
spin $S$ and relative orbital angular momentum $L$ of the final states
coupled to the $1^{--}$  $\psi$  state, which are  $SL=01, 21$, and
$23$. The cross-section data of three-body $D\bar D\pi$ channel is
assumed to be predominantly contributed by the two-body $D\bar D^*_2$
mode, as demonstrated in Belle's analysis~\cite{Belle:2007xvy}.
Therefore, the $D^0\bar D^-\pi^+$ data could be converted to $D\bar
D^*_2$ data using the branching ratio of
$\frac{\mathcal{B}(D^*_2\rightarrow
D\pi)}{\mathcal{B}(D^*_2\rightarrow D\pi)+\mathcal{B}(D^*_2\rightarrow
D^*\pi)}=0.62$~\cite{ParticleDataGroup:2022pth}. Thus,  by counting
the different charged states and  particle and antiparticle states as
different channels,
we have included totally 16 coupled hadronic channels~(as the
continuum states in the GFL model) with different particle species and
$SL$ quantum numbers, which could be categorized into six kinds:
$D\bar D$, $D\bar D^*$, $D^*\bar D^*(SJ=01)$, $D^*\bar D^*(SJ=21)$,
$D^*\bar D^*(SJ=23)$, $D\bar D_2$, in which three $D^*\bar D^*$
channels are degenerate and could be regarded as one channel. We have
not included the other channels such as $D_s^{(*)}\bar D_s^{(*)}$, $D\bar
D_1$, and hidden-charm channels in this analysis. Since in the QPC model, it is
more difficult to produce $s\bar s$ from the vacuum than the $u\bar u$
and $d\bar d$ because of the heavier $s$ quark mass, and the
OZI-suppressed channels only couple weakly to the resonance, we
would expect that omitting this channels does not influence of
our conclusion much. Besides, including all these channels will
introduce more free parameters and will
significantly increase the computational complexity which exceeds
our present computing capacity.

\begin{figure}[t]%
\begin{center}%
\includegraphics[height=28mm]{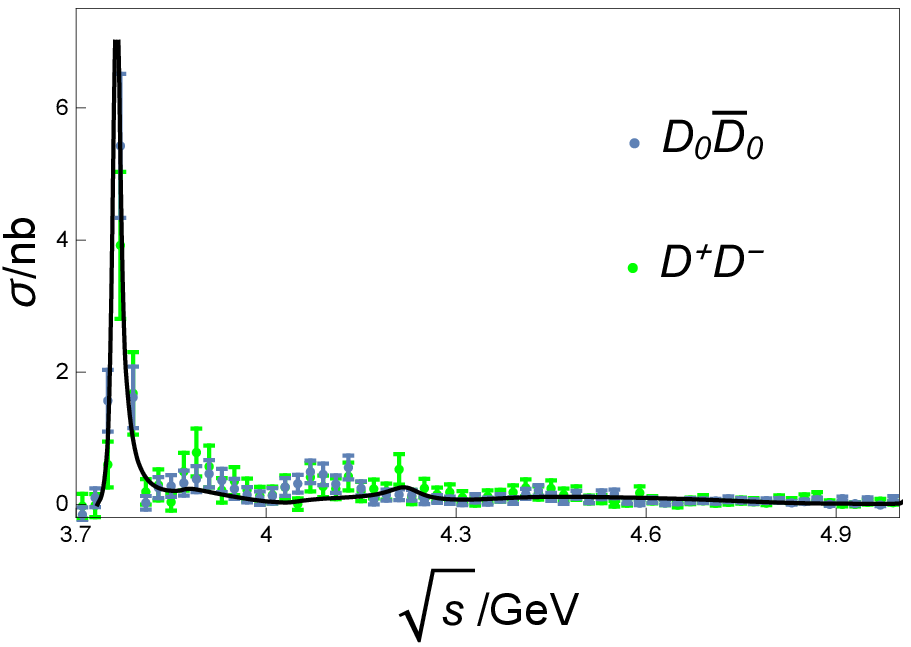}
\includegraphics[height=28mm]{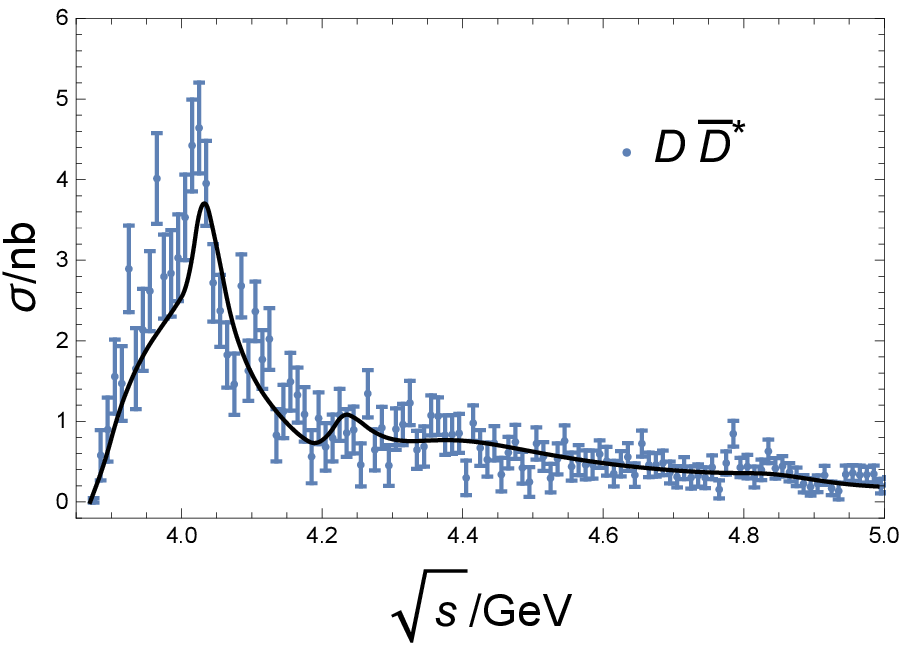}
\includegraphics[height=28mm]{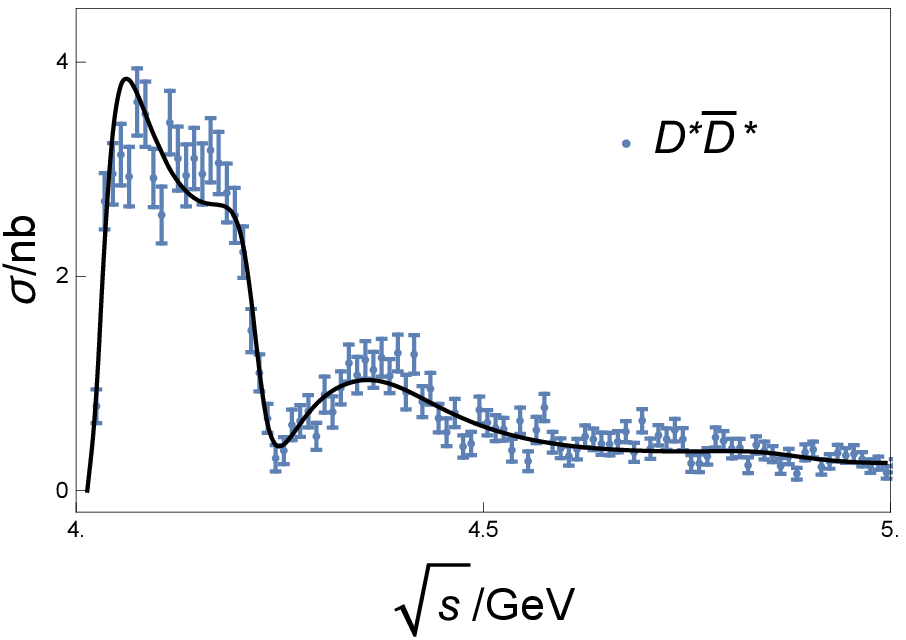}
\includegraphics[height=28mm]{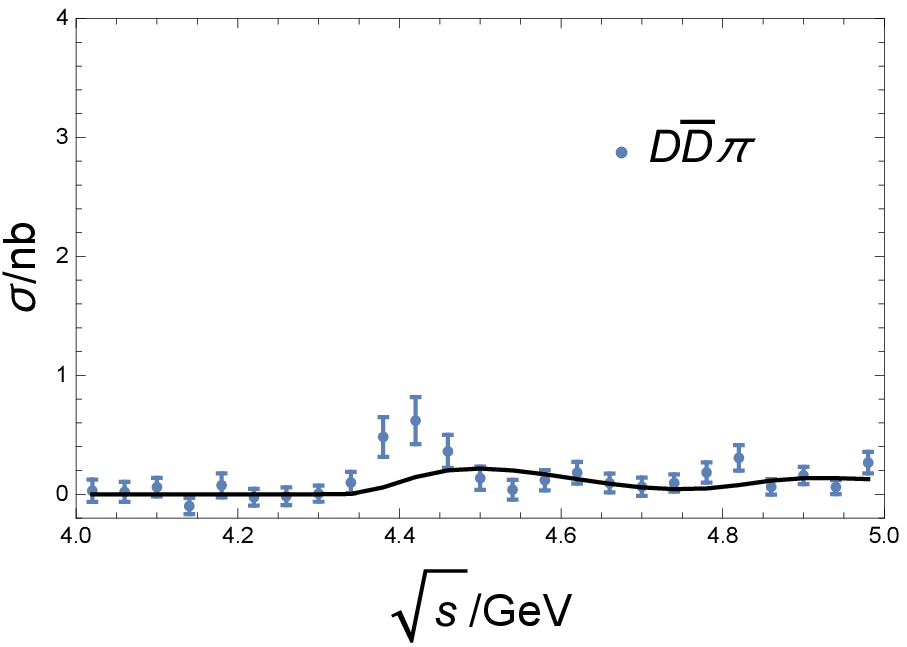}
\caption{\label{fitresult} The fit results of the cross sections of $e^+e^-\rightarrow D^{(*)}\bar D^{(*)}$ and $D\bar D\pi$ by Belle.}
\end{center}%
\end{figure}%

With regard to the discrete $\psi$ states, we have not made any
presumptions regarding the existence of additional bare $1^{--}$
$\psi$ states with exotic origins in the energy
range. Within the energy range of $E=3.73\sim 5.00$ GeV where the data
is collected,  only conventional $\psi(2^3S_1)$,
$\psi(1^3D_1)$, $\psi(3^3S_1)$, $\psi(2^3D_1)$, and
$\psi(4^3S_1)$  are included in the GFL
model, which may influence the cross section.
We employ the QPC model, which utilizes the wave functions that are
entirely determined by the quark potential model developed by Godfrey
and Isgur (GI)~\cite{Godfrey:1985xj}, to calculate the coupling
functions $f_{\lambda i}$ between the discrete state $\psi_\lambda$
and the continuum state denoted by $i$ from the
16 two-body states mentioned above. Hence, the only
parameter that remains undeterminate is the global quark pair
production strength from the vacuum, designated as $\gamma$ in
ref.~\cite{Zhou:2017txt}. However, we do not anticipate that the mass
of the bare $\psi$ states in GI's model is precisely calibrated to
reproduce the cross section data and hence treat them as free
parameters.

The coupling of the the $e^+e^-$ to the  $\psi$ states and their
relative phase could not be fixed by the model, so they are
also left free and to be determined by the fit. Therefore, there are
totally 15 free parameters to be determined in the fit: the QPC strength $\gamma$,  the bare masses of the five $\psi$ states,
their coupling constants  to the $e^+e^-$ states $g_{e\rho}$ and
relative phases $\phi_\rho$  (with one global phase undetermined).
\begin{figure}[t]%
\begin{center}%
\includegraphics[height=40mm]{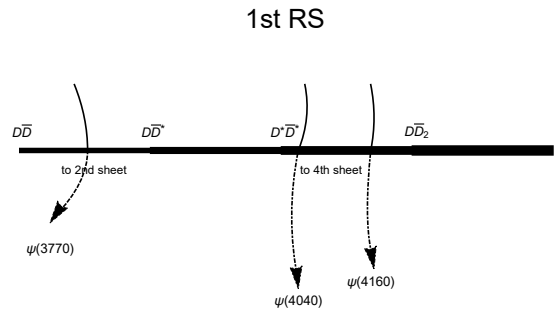}
\caption{\label{RS} A schematic depiction of the thresholds and the unphysical RSs in which the related poles are located. The dashed line means entering the second RS and the dashdotted line means entering the 4th RS.}
\end{center}%
\end{figure}%

The quality of the best fit with $\chi^2/d.o.f.=379/(293-15)=1.36$
is illustrated in Fig.~\ref{fitresult}.  It can be seen that the cross section data in the energy range of 3.73 to 4.3 GeV are well-described within the current theoretical
framework. Of particular significance is are the extracted nearby pole
positions located in the RS attached to the physical region,  as illustrated in Fig.~\ref{RS}. There is one second-sheet pole and two fourth-sheet poles, whose positions are as follows
\bqa
&z_0^{II}&=3.762-\frac{0.020}{2} i\mathrm{GeV}, z_{0,1}^{IV}=4.028-\frac{0.068}{2}i\mathrm{GeV}, \nonumber\\
&z_{0,2}^{IV}&=4.222-\frac{0.064}{2}i\mathrm{GeV}.
\eqa

\begin{figure}[t]%
\begin{center}%
\includegraphics[height=28mm]{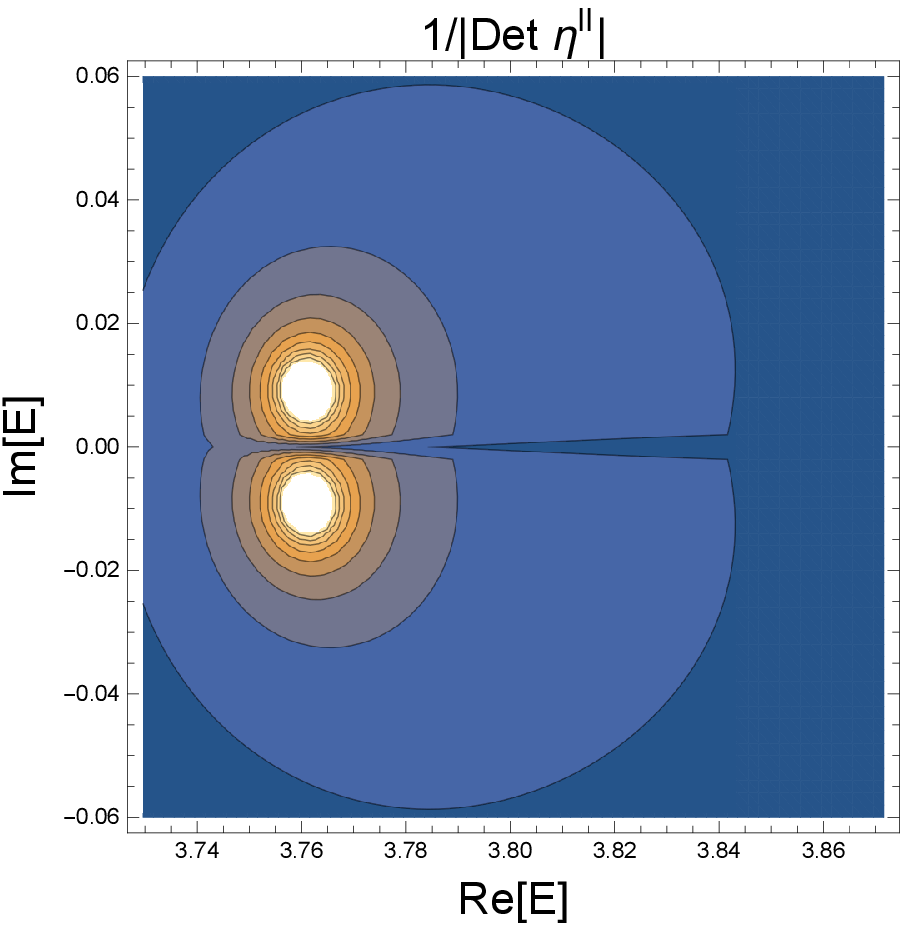}
\includegraphics[height=28mm]{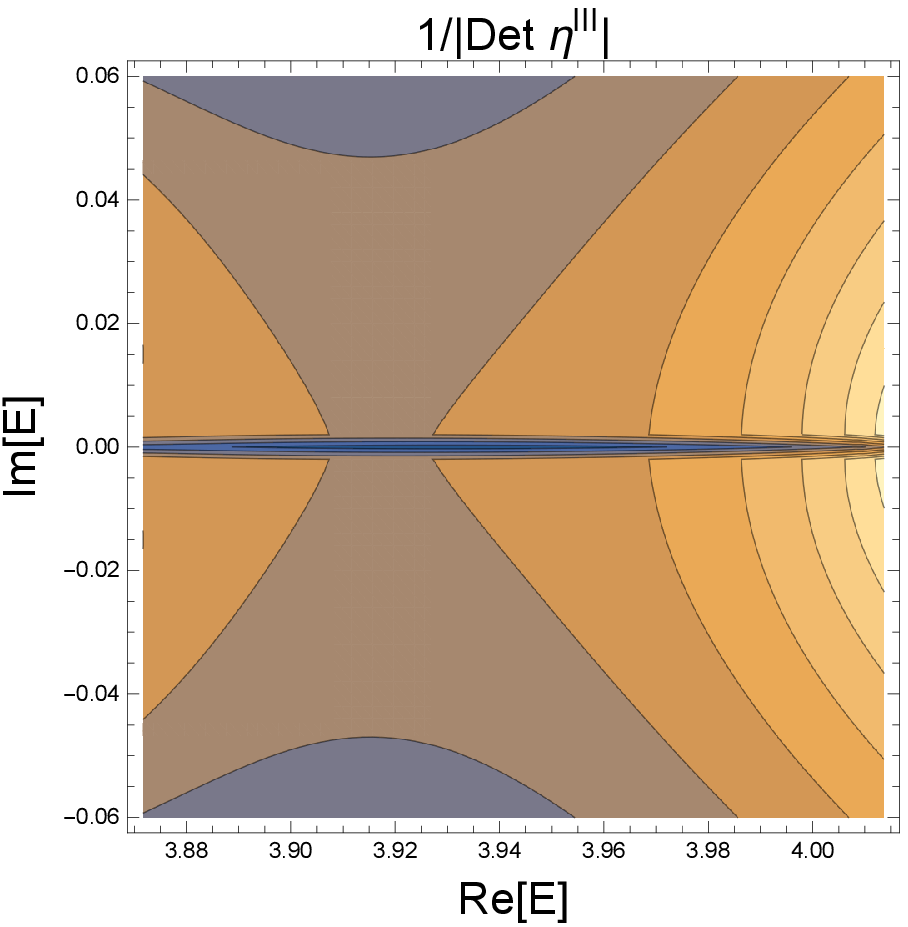}
\includegraphics[height=28mm]{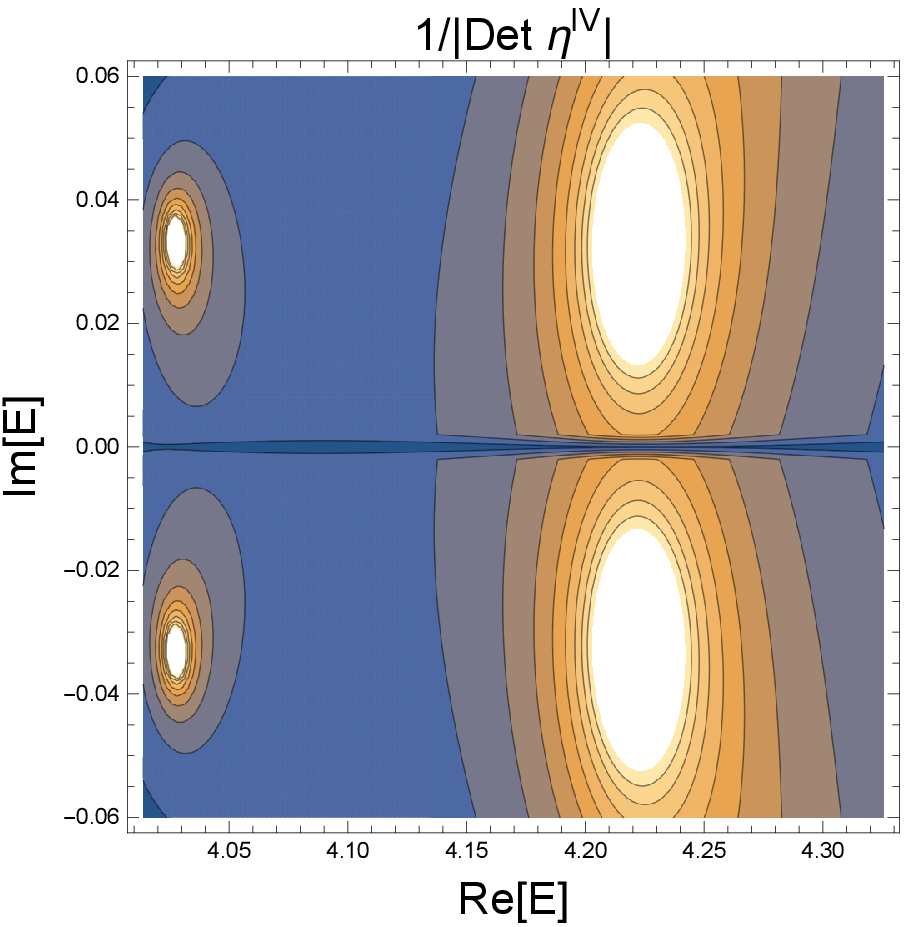}
\caption{\label{poles} The contour plots of $|1/\mathrm{Det}\eta|$ in the 2nd~(left), 3rd~(middle) and 4th~(right) Riemann sheets close to the physical region. The locations of resonance poles are shown in the complex energy planes. The units are GeV.}
\end{center}%
\end{figure}

The cut structures and the related poles are
depicted in Fig.~\ref{RS}, and the pole locations illustrated by
$1/|\mathrm{Det[\eta^{N}]}|$ are shown in Fig.~\ref{poles}.  In this fit, there is no third-sheet pole with its mass lying between the $D\bar D^*$ and $D^*\bar D^*$ thresholds.
It is noteworthy that  the pole information is
solely dependent on the bare mass parameters and $\gamma$,
the quark pair creation strength from the vacuum, and is
independent of the electroweak interaction coefficients.

The pole masses and widths of $z_0^{II}$ and $z_{0,1}^{IV}$  are in
agreement with the values of  $\psi(3770)$ and $\psi(4040)$ respectively as listed in PDG
table, which are conventionally determined using the Breit-Wigner
parameterization.

\begin{figure}[t]%
\begin{center}%
\includegraphics[height=40mm]{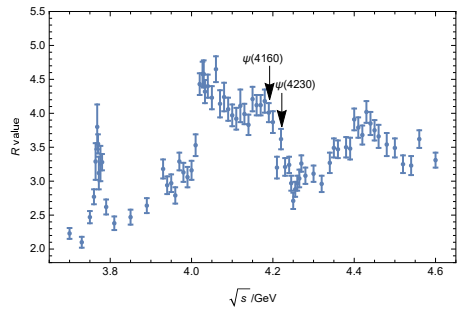}
\caption{\label{Rvalue} The $R$ values in the range of $3.7\sim 4.7$ GeV measured by BES~\cite{BES:2001ckj}. For comparison, the arrows indicate where the central masses of $\psi(4160)$ and $\psi(4230)$ listed in the PDG table~\cite{ParticleDataGroup:2022pth}.}
\end{center}%
\end{figure}

Nonetheless, the pole location for $\psi(4160)$ in this study deviates
from the value obtained from the fit with the BW parameterization,
exhibiting a higher mass. Rather than being
located at the
cross-section shoulder of $e^+e^-\rightarrow D^*\bar D^*$, it is found
at the half hillside near $4.22$ GeV.  It is  a natural outcome when a
resonance exhibits a substantial interference effect  with the
background, akin to the
example  shown in the
textbook \cite{Taylor:1972pty}. %page 242
It is
instructive to notice that  the pole mass and
width of  $\psi(2^3D_1)$ is similar to the average ones of  $\psi(4230)$
quoted in the PDG table~\cite{ParticleDataGroup:2022pth}
\bqa
M= 4222.7\pm 2.6 \mathrm{MeV},\Gamma=49\pm 8 \mathrm{MeV}.
\eqa
This observation implies that we may not need to adopt two different
states here and assign an exotic origin to one of them. Or, it at
least means that, interfering with other resonances,  a pole at about
$E=4222-32i$MeV could reproduce the sharp decline in the cross section
of $e^+e^-\rightarrow D^*\bar D^*$ near 4.22GeV.
It is worth pointing out that the
experimentally measured $R$ values, $R=\frac{\sigma(e^+e^-\rightarrow
hadrons)}{\sigma(e^+e^-\rightarrow \mu^+\mu^-)}$ also exhibit a similar sharp
decline at about 4.2 GeV as shown in Fig.\ref{Rvalue}, since it is the sum of all the exclusive
processes. This may also be a hint that $\psi(4160)$
and $\psi(4230)$ could be a same state. Moreover, the $\psi(4160)$ and $\psi(4230)$ being the same state
might provide a much simpler explanation for the difference of their
enigmatic open-charm decay mode and hidden-charm modes.

There is some more remarks about the fit results.  The fit parameters, as depicted
in Table \ref{fitparameters}, merit attention, as they are not
directly indicative of resonance information. Given the coupling to
the continuum, the  pole on the real energy axis is often shifted
downward from the bare mass to the complex energy plane, with the
magnitude of the shift depending upon the
coupling model. In this
instance, the $\psi(2S)$ is not well described. One of the reason might be  that the
data points beginning at the $D\bar D$ threshold,  which is above  the
 mass of $\psi(2S)$  and thus provides little information to fix
$\psi(2S)$.  As a result,
it plays a role of an overall background absorbing all the other $1^{--}$ states
and background  not incorporated in this analysis.
%Additionally, the extensive
%uncertainty of the data set in  $D\bar D\pi$
%channel also  poses a challenge for
%the global fit capacity to accurately represent the higher
%$\psi(4415)$ resonance with only a few free parameters.

\begin{table}[b]
\begin{center}
\begin{tabular}{|c|c|c|c|c|c|}
  \hline
     &$\psi(2S)$&$\psi(1D)$&$\psi(3S)$&$\psi(2D)$&$\psi(4S)$\\
\hline
Bare mass &1.83 & 4.09 & 4.36 & 4.62 & 5.35 \\
\hline
   $g_{e\rho}$ &26.4 &1.82 &5.32 & -3.68& -6.12 \\
\hline
   $\phi_{e\rho}/^\circ$ & 0(fixed) & 35.2 & -31.6 & 64.6& 129.3 \\
\hline
  $\gamma$ &\multicolumn{5}{c|}{4.48 } \\
  \hline
\end{tabular}
\end{center}
\caption{\label{fitparameters}The results of 15 fit parameters. The unit of bare mass is GeV.}
\end{table}%

We wish to underscore that the present framework is formulated in a
non-relativistic context, as it is considered as a suitable approximation
for the hadronic states associated with open-charm channels. Governed
by the QPC model and the GI wave function, the interaction and mixing
of these states put high restriction on the analysis. This constraint on
the freedom of tuning the parameters renders it challenging to achieve
compatibility with all available data. In the current fitting process,
the inability to identify a viable solution  of  the
$\psi(4415)$ may be attributed to the considerable uncertainty
associated with the $D\bar D\pi$ data. A more intricate relativistic
variant of the Friedrichs-Lee scheme has been
established~\cite{Xiao:2023}, and a comprehensive coupled-channel
analysis encompassing a larger number of channels and free parameters
for fitting the $R$ value data and other inclusive processes could
potentially yield a more conclusive outcome.

The present study  is distinct from
ref.~\cite{Uglov:2016orr}  primarily  because of its analytical
structure and the derivation of coupling vertex factors. In contrast,
ref.~\cite{Uglov:2016orr} is primarily concerned with describing the
lineshapes of cross-section data. It is noteworthy that the current
analysis's emphasis on developing an analytical framework to
comprehend particle coupling mechanisms sets it apart from the
previous study. It is the analytical structure of the amplitude that
leads to extracting the pole position of $\psi(4160)$  that can not
be analyzed in ref.~\cite{Uglov:2016orr}.

\emph{Summary}:   For the first time,  based on a coupled-channel
unitary scheme, we have
extracted the pole position of $\psi(4160)$  by
analyzing
 the cross section
data of  $e^+e^-\rightarrow D^{(*)}\bar{D}^{(*)}, D\bar D\pi$
processes. The result indicates
that the mass of $\psi(4160)$ might be higher than the value commonly
 accepted  and close to the mass of  $1^{--}$ $\psi(4230)$ state. The
pole of $\psi(4160)$ being located at the half hillside of the cross section
data,  though being counterintuitive,
could
be  a result of a significant interference effect, which may lead to
a strong hint that  $\psi(4160)$ and $\psi(4230)$ could be the same
state, $c\bar c$ $2^3D_1$ mixed with continuum. Further analysis along
these lines on
the other experimental data such as $R$ values, and $e^+e^-\to
J/\psi\pi\pi$ etc. need to be done to confirm this suggestion.  If this  result is proved by other
analyses,  it could be of great help to understand the
enigmatic decay properties of the $\psi(4160)$ and $\psi(4230)$ listed
in the PDG table.

\textit{Acknowledgement:} Helpful discussions with  Dian-Yong Chen and Hai-Qing Zhou are appreciated. This work is supported by China National Natural Science Foundation under contract No. 11975075, No. 11575177, and No.11947301.

%\bibliographystyle{apsrev4-1}
\bibliography{Ref}
\bibliographystyle{apsrev}

\end{document}